# DEFINING COMPLEXITY

## A Commentary to a paper by Charles H. Bennett

*By Mark Perakh*

This is a commentary to an article by Charles H. Bennett, titled "How to Define Complexity in Physics And Why" which is found in the anthology *From Complexity To Life*[1] . Originally this commentary was meant to be a section in the review of the entire anthology in question. I started reviewing various papers in that anthology in the fall of 2003 but was distracted by other projects. Then Norman Levitt wrote an excellent review of the anthology in question thus significantly devaluating the incentive for writing one more review of the same anthology. At the time Levitt's review appeared, I had the reviews of three papers in that anthology – those by Bennett, Davis, and Stewart - partially prepared.

Although I did not completely abandon the idea of finishing this project, it was relegated to a low priority shelf. It rested on my disk waiting its turn, whenever its time might come.

Then something happened that prodded me to pull the half-finished review of Bennett's paper from the disk and take a look at it. The event that caused this renewed interest in Bennett's paper was the recent appearance of a paper by William Dembski (see http://www.iscid.org/ubb/ultimatebb.php?ubb=get_topic;f=6;t=000533 and of its critique by Cosma Rohilla Shalizi (see
http://www.cscs.umich.edu/~crshalizi/weblog/234.html ).

As Shalizi has shown, Dembski has demonstrated in his paper an amazing lack of familiarity with the literature in the field of which he has been acclaimed an expert. In particular, the quantity Dembski introduced in his new article and named Variational Information, is in fact a well known for more than 40 years Rènyi divergence of the second order.

Dembski reaction to Shalizi's essay was immediate but evasive. In his response, Dembski names me several times and takes the liberty of asserting that I, as well as a number of other critics (Shallit, Levitt, Wein, Schneider, and Stenger) have no knowledge of the sophisticated math he used in his paper and that is why we turned to Shalizi for help (as he put it, "Perakh is out and Shalizi is in.").

I can't speak for Shallit, Levitt, Wein, Shneider, or Stenger. I am confident, though, that at least Shallit, Leviit, and Wein, who are mathematicians, must be better versed in the subject matter of Dembski's paper than I, a physicist rather than a mathematician. Nevertheless, even I had no problem of understanding Dembski's paper, and form an opinion of its merits, while the above listed mathematicians were certainly in even a better position to judge it.

Still, perhaps indeed none of these mathematicians (and even less I) are experts in the specific subfield of mathematics to which Dembski's paper purports to belong. What is wrong with that? Likewise, there are areas of knowledge where I or anyone of the listed colleagues may be well versed while Dembski lacks any knowledge. The difference is, though, that neither I, nor the above listed colleagues pretend to be experts in the fields beyond our specialization, and even in the fields of our expertise we never ourselves claim to have suggested revolutionary breakthroughs and, unlike Dembski, are not compared by our colleagues to Isaac Newton.

It was only naturally that a real expert in the field in question, such as Shalizi, undertook the task of dissecting Dembski's paper, and did it in the most convincing way.

Moreover, Dembski has no way to know what the actual level of my or of those mathematicians' knowledge in the field in question is. Therefore his assertion that the mathematical apparatus he used in his new paper is well over our heads is not based on any factual information but is rather a display of his arrogance.

My review of Bennett's paper has no relation whatsoever to Dembski's new article and to his allegedly innovative approach to information measures. However, it happens to contain references to the works of Rènyi and of other researchers who utilized and further developed Rènyi's ideas.

Having encountered the arrogant comments of Dembski according to which I am not qualified to judge his great breakthroughs in information theory, I decided to separate my old unpublished review of Bennett's paper from the rest of the planned review of the entire anthology by Gregersen and to post it now as a stand-alone piece, and in this way to demonstrate that I, not being a mathematician, am perhaps still better versed in the literature on information theory than the acclaimed expert Dembski.

By no means I claim to be an expert in this field. My knowledge of Rènyi's theories is limited and rather shallow. Until Shalizi has pointed to the fact that Dembski's Variational Information is nothing more than Rènyi divergence of the second order, I myself did not realize that. However, after Shalizi posted his critique, I had no difficulty to see the validity of his comments. Dembski's arrogant remark that I and other critics of his work lack qualification to judge his paper does not sound convincing given his own insufficient familiarity with the literature in the field where he is supposed to offer important new results.

Here starts my review of Bennett's paper.

Bennnet's paper is a survey of various definitions of complexity, and specifically of their utility for physicists. This is a hot point in modern science and there is a vast literature devoted to the elucidation of the concept of complexity.

Charles H. Bennett is a Senior Scientist at IBM's Thomas J. Watson Research Center in Yorktown Heights, NY. He is a co-inventor of quantum cryptography and is credited with other important contributions to science. While I hold Bennett in a high esteem, in my view this particular paper is deficient in some respects.

Bennett suggests a classification of possible definitions of complexity. To my mind his classification is sometimes based on vague criteria, and incomplete.

Strangely, Bennett omits mentioning some of the definitions of complexity which have been discussed recently (but prior to the 1999 conference at the Santa Fe institute where the articles collected in this anthology were presented). An example of such an omission is "LMC complexity" suggested in 1995 by Lòpez-Ruiz, Mancini, and Calbet.[2]

LMC complexity is one of "statistical complexities" (as distinctive from "deterministic" complexities, such as Solomonoff-Kolmogorov-Chaitin complexity, more often referred to as simply Kolmogorov complexity).

Perhaps Bennett does not view LMC complexity as a fruitful concept, but, in my view, this does not justify ignoring it in an article devoted to the survey of definitions of complexity.

Here is a brief explanation of LMC complexity. It was suggested as the product of two quantities – Shannon's entropy (which is equivalent to Boltzmann-Gibbs entropy) and a quantity named disequilibrium. The latter is a function reflecting in a rather simple and transparent way the divergence of a given probability distribution from uniformity (see the Appendix to this review). This product of entropy and disequilibrium vanishes in the two extremes – perfect order and perfect disorder, which is a necessary (even if perhaps not sufficient) property of a complexity measure.

Perhaps Bennett shares the view of some critics of the LMC concept, such as Feldman and Crutchfield.[3] These authors have shown that LMC complexity is neither an intensive nor an extensive quantity in the thermodynamic sense. To remedy this "weakness" of LMC complexity, Feldman and Crutchfield suggested a modified LMC complexity wherein one of its component – the disequilibrium – is replaced by the "relative information" also known as Kullback-Leibler divergence (or as Rènyi divergence of the first order – see, for example [4]). This substitution converts LMC complexity into an extensive quantity (in the sense that it increases linearly with the system's size). However, as Feldman and Crutchfield conclude, the modified LMC complexity is in fact a trivial function of entropy density and as such is of a limited use as a measure of structure or memory.

Since Bennett did not mention either LMC complexity or its critique by Feldman and Crutchfield, the reader is left in dark as regards Bennett's view on that matter and the place of the concept in question among other definitions of complexity.

Here is my two pence regarding LMC complexity and its critique. To my mind, LMC complexity has its legitimate place as one of the statistical definitions of complexity, despite its shortcomings. Feldman and Crutchfield's analysis is well taken and, indeed, LMC complexity is not a linear function of the system's size, as the conventionally defined extensive quantities usually are. However, it is not at all clear to me that this is such a serious shortcoming. LMC complexity is still increasing with the system's size (which testifies to its being not an intensive quantity), albeit not linearly. Perhaps, it can be construed as a quantity extensive in a more convoluted manner than the more conventional extensive quantities such as, for example, mass.

Furthermore, perhaps LMC complexity can indeed be gainfully modified in the spirit of Feldman and Crutchfield's amendment without making it a trivial function of entropy density. One way to do so is possibly using instead of Kullback-Leibler divergence, which is just the first-order Rènyi divergence, Rènyi divergence(s) of higher order (perhaps the second order version is a suitable candidate). In doing so, we may expect to lose again the linear dependence of the measure in question on the system's size, but some of LMC-like complexities using higher order Rènyi divergences instead of either disequilibrium or Kullback-Leibler divergence may open interesting new ways of estimating complexity.

Another shortcoming of LMC complexity, in Feldman et al.'s view is that it is "over-universal". This term means that this quantity, while reflecting the system's "disorder," may have the same value for structurally different system. Feldman et al. think that such property of the quantity which is supposedly measuring complexity is a serious weakness. In my view, while Feldman et al.'s objection, again, is well taken, the ultimate conclusion regarding the utility of LMC complexity can be made only after it has been applied to specific problems and its behavior in various practically relevant situations has been observed.

I think it is possible that LMC complexity, despite its imperfection, can presumably have some gainful use, especially given the fact that there is no perfect definition of complexity and no universal concept of complexity applicable to all situations. LMC complexity has an advantage of being a simple concept amenable to easy interpretation and use in various problems. Indeed, there are a number of published papers wherein applications of LMC complexity are demonstrated and some of them look promising. [5]

I am not a champion of LMC complexity which may or may not become a useful even if only limited tool in the complexity theory – the champion's role legitimately belongs to the originators of that concept. My point here is not so much a defense of LMC complexity, as rather giving an example of Bennett's omitting discussions of some of the complexity measures suggested in literature.

Another definition of complexity absent from Bennett's paper is that by Shiner, Davison, and Landsberg.[6] These authors approached their task in a way similar to that by Lòpez-Ruiz et al., but suggested a different mathematical expression for statistical complexity. It is a quadratic function containing a quantity they call "disorder" (see the Appendix).  Crutchfield, Feldman, and Shalizi criticized Shiner et al.'s definition[7] as being, similarly to LMC complexity, "over-universal." (Bennett could not know of Crutchfield et al.'s critique which appeared after the conference at the Santa Fe institute. He possibly could have known of Shiner et al.'s paper which appeared in 1999.)

One more concept of complexity which Bennett did not mention, is that suggested by W. Dembski (despite Dembski's being one of the participants in the Santa Fe institute conference and a contributor to the anthology under discussion). Unlike LMC complexity, which may or may not be very useful but which entails certain reasonable notions, Dembski's complexity is, in my opinion, a concept making little sense. In fact, in different parts of his publications, Dembski uses the term complexity in different ways often incompatible with each other. [8,9]

Bennett endeavors to find a definition of complexity that would work in physics. Before embarking on that discourse, Bennett formulates his goal, which is to find definitions of complexity "that on the one hand adequately capture intuitive notions of complexity and on the other hand are sufficiently objective and mathematical to prove theorems about." This certainly is a very interesting and worthy endeavor which, if successful, would enable important theoretical and practical applications. Bennett's preferred approach to the definitions in question is to favor the notion of "logical depth," of which he, I believe, was the originator. To explicate this notion, Bennett refers to his other publications. Since I am discussing here only the collection *From Complexity to Life*, I will not comment on Bennett's other publications.

A reader not familiar with Bennett's other publications will remain uncertain in regard to the meaning of logical depth and its relation to the notion of complexity. (The logical depth of, for example, a system's configuration, is the time required for a universal Turing machine to run the minimal program that reproduces it. Obviously, this concept requires a more detailed elucidation for non-experts, but Bennett unfortunately does not provide such, so his discussion of logical depth will leave many readers as much in dark in that respect as before reading his paper.)

Before going into details of his concepts of complexity, Bennett observes, "Life-like properties (e.g., growth, reproduction, adaptation) are very hard to define rigorously, and are too dependent on function, as opposed to structure." This observation seems almost self-evident, but to my mind it is unclear why it should exclude a definition of complexity based on structure. Bennett's example – "a dead human body is still complex, though it is functionally inert" seems to jibe with the idea that complexity can be fruitfully defined by accounting only for the properties of the structure and divorced from the consideration of function. Functionality and complexity are two different concepts and in my view there is no need to make functionality interfere with the *definition* of complexity. As long as our goal is limited only to finding a workable definition of complexity, the discussion of the relation between complexity and functionality is better to be postponed until the next step of discourse. The high complexity of living organisms may be analyzed as a separate phenomenon, and its relation to functionality can be discussed on a subsequent step apart from the *definition* of complexity.

Bennett proceeds by pointing out that thermodynamic potentials (he lists only two, entropy and generic free energy) are unsuitable as measures of complexity. To me this statement looks self-evident and I can't recall the thermodynamic potentials (including all those not listed by Bennett) ever suggested as measures of complexity (except, perhaps, some forms of entropy[2] which, however have not been really suggested as profitable measures of complexity in a serious way). From the above trivial observation Bennett derives a supposedly new law labeled the "slow growth law." The essence of that putative law is that "complexity ought not to increase quickly, except with low probability, but can increase slowly, for example, over geological time."

Unfortunately, besides the vagueness of the suggested law which contains no indications what the terms "quickly" and "slow" mean quantitatively, Bennett provides no arguments supporting the validity of the new law. His supporting notions boil down to a couple of examples which can have various interpretations.

One of his examples is about phase transformation – a rapid "crystallization of a supersaturated solution following introduction of a seed crystal." This process is well understood and explained, for example, in thermodynamics. In this process, the first and the second laws of thermodynamics are sufficient to account for the process's features (although its details may be better elucidated using thermodynamic potentials, and even more features can be revealed by applying physical kinetics). Bennett compares this process with "rapid growth of bacteria following introduction of a seed bacterium" into a sterile nutrient solution. In Bennett's view, the growth of bacteria is a process which is not a manifestation of the second law of thermodynamics, but rather of a new law of "slow growth." To my mind, all this discourse is obscure and somehow lacking a sufficient logical or empirical substantiation. I don't think any scientist would ever maintain that the growth of bacteria is the same process as the crystallization of a supersaturated solution, so Bennett seems to state the obvious. However, how the difference between the crystallization of a supersaturated solution and the growth of bacteria leads to a putative "law of slow growth" is, in my opinion, not really explained. If there is a logical connection here, Bennett should have explained it instead of simply claiming such a connection. As the matter stands now, the real judgment on the suggested new law has, in my view, to wait until Bennett provides lucid arguments in favor of that law.

Bennett continues by discussing such variants of complexity as computational universality, computational time/space complexity and algorithmic information. I will leave without comments the first two items as I have no objections to Bennett's discussion in these two sections, but I'll briefly comment on the third. Bennett writes, "…algorithmic entropy corresponds intuitively to randomness rather than to complexity." This sounds mysterious because algorithmic complexity (often referred to as Kolmogorov complexity[9] ), within the framework of Kolmogorov-Chaitin's theory is to all intents and purposes tantamount to randomness: the more random a string the larger its Kolmogorov complexity which, as Bennett correctly states, is the size of the shortest program generating the string in question. While algorithmic theory by Kolmogorov –Chaitin is not perfect (for example, because it is impossible to assert randomness of a finite string, or because the shortest program can only be defined allowing for a certain fudge factor) it nevertheless is theoretically powerful in the domain of its applicability, even if not always useful practically. Bennett concludes this section with a correct observation that "…complex genome or literary text is intermediate in algorithmic entropy between a random sequence and a prefectly (sic) orderly one." He seems to view this as a drawback of the concept of algorithmic entropy. To my mind, it is not really a drawback. I discussed a similar notion in detail (see for example[10]).

This assertion does though work against a sometimes suggested notion that informational entropy is a negative of information. I believe such a notion is misleading (a more detailed discussion of this point will be offered in an essay, which is in preparation, about the paper by Ian Stewart[11] in the same anthology).

NOTES
1. Charles H. Bennettt. "How to Define Complexity in Physics and Why." In Niels Henrik Gregersen, editor, *From Complexity To Life: On the Emergence of Life And Meaning*. Oxford University Press, New York, 2003.
2. R. Lòpez-Ruiz, H. L. Mancini, and X. Calbet. "A Statistical Measure of Complexity." Phys. Lett., A209, 321-326 (1995). See also Ricardo Lòpez-Ruiz, "Shannon Information, LMC complexity and Rènyi entropies: a straightforward approach." http://arxiv.org/PS_cache/nlin/pdf/0312/0312056.pdf , Dec 22, 2003.
3. David P. Feldman and James P. Crutchfield, "Measures of Statistical Complexity: Why?" www.santafe.edu/~cmg/papers/mscw.pdf , November 11, 1997.
4. Alfrèd Renyi, *Probability Theory,* Amsterdam, North Holland, 1970. Also see Afrèd Rènyi. "On Measures of Entropy And Information." In *Proc. 4th Berkeley Symp. Mathematical Statistics Probability,* Berkeley, CA, Univ. Calif. Press, 1961, 547-561. For Kullback-Leibler divergence, see, for example, Z. Rached, F. Alajaji, and L.L. Campbell, "Rènyi's Divergence and Entropy Rates for Finite Alphabet Markov Sources." *IEEE Transactions On Information Theory,* v. 47, No 4, May 2001, pp 1553-1560.
5. X. Calbet and R. Lòpez-Ruiz, "Tendency Towards Maximum Complexity in an Isolated Non-equilibrium System." Phys. Rev., E6066116(9), 2001. Also G. Feng,

APPENDIX

The term "LMC complexity" was composed by its authors from the initials of their surnames: Lòpez-Ruiz, Mancini, and Calbet . The formula given by these authors is:

$$C = H \times D,$$

where C is complexity, H is Shannon entropy (which is an equivalent of Boltzman-Gibbs's entropy) and D is called disequilibrium. This quantity is a measure of the divergence of the given probability distribution from the uniform one.

$$H = -K \sum_{i=1}^{N} p_i \log p_i$$

$$D = \sum_{i=1}^{N} (p_i - \frac{1}{N})^2$$

Here $p_i$ is probability whose (finite) distribution runs over N values, and K is a constant depending on the choice of units (essentially entropy is a dimensionless quantity, so taking K=1 does not impair the general behavior of that quantity).

For the case of perfect order H vanishes, and for the case of complete disorder D vanishes, hence C vanishes in both extremes which is a necessary (albeit perhaps not a sufficient) requirement for a complexity measure.

Feldman and Crutchfield's amendment replaces D with the Kullback-Leibler divergence (which is the first order Rènyi divergence) which these authors also denote D. To avoid confusion, here it will be denoted $D_{R1}$ where the subscript R1 refers to it being the first order Rènyi divergence.

$$D_{R1} = \sum_{i=1}^{N} p_i \log \frac{p_i}{q_i}$$

where $p_i$ and $q_i$ are probabilities belonging to two different distributions; $D_{R1}$ "measures" the divergence between these two distributions. For the purpose of serving as a component of complexity, one of the compared distributions is taken to be uniform, which leads to

$$D_{FC} = \sum_{i=1}^{N} p_i \log N p_i$$

where the subscript FC refers to Feldman and Crutchfield.

If, instead of Kullback-Leibler divergence, a higher order Rènyi divergence were to be used, the expression for $D_{R1}$ or $D_{FC}$ would be replaced with

$$D_{R\alpha} = \frac{1}{\alpha - 1} \log(\sum_{i=1}^{N} p_i^{\alpha} q_i^{1-\alpha})$$

where $\alpha$ is the order of Rènyi divergence. In the limit of $\alpha$ approaching 1 it converts into Kullback – Leibler divergence shown above.

The above equations can be generalized for continuous distributions (where the sums are replaced with appropriate integrals).

Shiner-Davison-Landsberg (SDL) complexity is expressed as:

$$\Gamma_{\alpha\beta} = \Delta^{\alpha}(1-\Delta)^{\beta}$$

where $\Delta$ is called "disorder" and is defined as

$$\Delta = S/S_{max}$$

S is the Boltzmann-Gibbs-Shannon entropy of the system and $S_{max}$ is the maximum possible entropy (corresponding to the random distribution). If both α>1 and β>1, the SDL complexity vanishes for both extremes – perfect order and perfect randomness, thus satisfying the requirement for a complexity measure (which, though, may be not necessarily sufficient).